%% file: main.tex
\documentclass[a4paper]{article}

\usepackage{INTERSPEECH2022}
\usepackage{array,multirow,graphicx}
\usepackage[table]{xcolor}
\usepackage{color}
\usepackage{xspace}
\usepackage{amsmath}
\DeclareMathOperator{\sinc}{sinc}

\newcommand{\sincnet}{SincNet\xspace}
\newcolumntype{P}[1]{>{\centering\arraybackslash}p{#1}}
\title{Low-Level Physiological Implications of\\ End-to-End Learning of Speech Recognition}
\name{Louise Coppieters de Gibson$^{1,2}$, Philip N. Garner$^1$}
\address{%
  $^1$Idiap Research Institute, Martigny, Switzerland,\\
  $^2$Ecole polytechnique f\'ed\'erale de Lausanne (EPFL), Switzerland}
\email{louise.coppieters@idiap.ch, pgarner@idiap.ch}

\begin{document}

\maketitle

\begin{abstract}
Current speech recognition architectures perform very well from the point of view of machine learning, hence user interaction.  This suggests that they are emulating the human biological system well.  We investigate whether the inference can be inverted to provide insights into that biological system; in particular the hearing mechanism.  Using \sincnet, we confirm that end-to-end systems do learn well known filterbank structures.
However, we also show that wider band-width filters are important in the learned structure.  Whilst some benefits can be gained by initialising both narrow and wide-band filters, physiological constraints suggest that such filters arise in mid-brain rather than the cochlea.  We show that standard machine learning architectures must be modified to allow this process to be emulated neurally.

\end{abstract}

\input{Introduction}
\input{Background}
\input{Basic_analysis}
\input{Wideband_filters}

\section{Conclusion}

E2E training of filterbanks for ASR leads to filters that resemble a standard filterbank.  However, wider bandwidth filters are learned too, and are important for good ASR performance.  
The central frequencies of the narrow-band filters tend to a mel spacing, regardless of the initialisation.  This confirms a well understood mechanism, suggesting that it exists in the biological system.  There appears to be an optimal number of filters --- around 30 to 40 --- that also correlates with acknowledged literature.

We suggest that wide-band filters are learned to distinguish voiced (harmonic, coherent) components from either background noise or unvoiced (aperiodic, stochastic) components.  In principle, a network should be able to learn such wide-band components by combining narrow-band ones.  
We argue that this capability is precluded by the (otherwise standard) ML architecture; in particular the phase will be lost by maxpooling.  
This argument is borne out by experiments showing that a structure retaining a more formal down-sampling mechanism can lead to better performance.

We are not aware of structures in the inner-ear or cochlea that can emulate physical wide-band filters.  However the phase information is retained in our current best understanding of cochlear operation, retaining also the possibility that such filters are emulated in mid-brain.  Proving this would be difficult, perhaps requiring some combination of selective stimulii with MRI or EEG sensing.  It remains as a hypothesis for the neuroscience community.
Our own future work will focus on more biologically plausible architectures.  This experimental study indicates that any such model will need to retain phase in order for the subsequent network to take advantage of both narrow-band and wide-band features.

\section{Acknowledgements}

This work was supported as a part of NCCR Evolving Language, a National Centre of Competence in Research, funded by the Swiss National Science Foundation (grant number 180888).

\bibliographystyle{IEEEtran}

\bibliography{references}

\end{document}

%% file: Introduction.tex
\section{Introduction}

Advances in automatic speech recognition (ASR) have led to performance that is very good in terms of word error rate (WER), but perhaps at the expense of our own understanding of how they function.  End-to-end (E2E) techniques \cite{amodei2016deep} have removed the need for knowledge of the hearing mechanism. Self-supervised training \cite{schneider2019wav2vec} has done the same for phonetics.  More generally, large pre-trained models are available \cite{babu2021xls} removing the need for even the machine learning (ML) know-how.  Given that these systems work well, the question arises: ``what have they learned?''  This is difficult to answer because their component parts cannot readily be mapped to biological ones.

In this study, we are interested in the hearing mechanism.  The biological mechanism is quite well understood \cite{Lyon2017}, with important parts being the logarithmic response to both frequency and amplitude.  For many years, filterbank approaches were used as models of this process \cite{juang2005automatic, hermansky1990perceptual}.  Whilst many variations have been studied, the authors' ad-hoc experience suggests that the details do not lead to big changes.  Recent E2E approaches, however, have clearly demonstrated that training the filterbanks can be beneficial \cite{collobert2016wav2letter}. A (1D) convolution layer in the machine learning field is a filter in the signal processing field.  However, the only validations of which we are aware tend to show that the component convolutions learn filters with a distribution similar to a mel filterbank.  This in turn tends to reinforce the above question rather than answer it.

In \sincnet, Ravanelli et al. \cite{ravanelli2018speaker} constrained the convolutions to be a sinc ($\sin(x) / x$) form, leading to a rectangular band-pass filter.  The filter is then defined by two trainable parameters: the lower frequency and the bandwidth.  Whilst not being biologically accurate, this approach has a distinct attraction of being interpretable.

In the remainder of this mainly experimental paper, we describe \sincnet and a modest frame-based experimental scenario.  We confirm that \sincnet learns a mel filterbank, but also show that wider bandwidth filters are important for performance.  We argue that such filters arise because of restrictions of standard ML convolutional architectures, and conclude with what this infers about how to construct a biologically plausible hearing model.

%% file: Background.tex
\section{Background}
The study of the human cochlea has interested many researchers since the beginning of the 20th century. Von Békésy laid the groundwork of the research on this topic in 1960 \cite{Bekesy1960}. 

The basilar membrane in the cochlea can be interpreted as a natural filterbank \cite{geisler1976mathematical, zwislocki1953review}.  Current understanding of the working of the cochlea is that wave propagation is an active process \cite{de1983active} and works as an array of Hopf oscillators \cite{hudspeth2008making, Hudspeth2010ACO}.  However, in this study, we limit ourselves to passive analogues. The scaling of this natural filterbank has been analysed from different points of view, leading to several scaling (or warping, spacing) approaches. 
The Greenwood scaling \cite{sridhar2006frequency} is the one that best represents the scaling of the frequency sensitive hair cells in the cochlea based on the physical distance on the basilar membrane of the hair cells.
The mel scale \cite{stevens1937scale} is based on frequencies judged to be equally spaced in human perceptual tests.  Bark \cite{Zwicker1961,smith1999bark} and ERB (equivalent rectangular bandwidth) \cite{moore1983suggested} are somewhere between mel and Greenwood, but by contrast are derived from critical \emph{bands} of hearing.


ASR frontends take either some preprocessed features as input or, more recently, raw input waveforms. 
Filterbanks have been the basis of feature extraction \cite{shannon2003comparative} for many years. 
As early as 2001, a study \cite{burget2001data} showed that a filterbank could be obtained from a mathematical derivation of a data driven design. From the resulting filterbank, about half of the filters were then kept to represent the filterbank motivated by the fact that those filters were enough to cover the whole frequency range.
For the E2E approaches, CNNs for ASR were introduced by Hinton et al. \cite{hinton2012deep, palaz2013estimating} and have been used for a decade.
Since 2018 some architectures propose a way to combine both the filterbanks and the E2E architecture, where the filterbanks are trainable and part of the convolution layers. Zeghidour et al. \cite{zeghidour2018learning} proposed an implementation with with Gabor filters and Ravanelli et al. \cite{ravanelli2018speech} with rectangular filters (SincNet). 
Other work on trainable filterbanks includes that of Seki et al. \cite{seki2019discriminative}, who proposed an architecture based on a filter layer combined with a DNN where the filter features were directly computed with a log-compression after the filter layer. 
In that study the gain, central frequency, bandwidth and filter shape were free to train, whilst in SincNet only two parameters are free to train, defining the filters in the first layer.


%% file: Basic_analysis.tex
\section{Initial Analysis}
In an initial, quite basic analysis, the main motivation was to understand what the trainable filters learn; i.e., which typical hyperparameters (e.g., the number of filters needed to describe the signal) can be derived from those trainable filter models, which initializations are appropriate.  For this study, we focus on SincNet.

\subsection{SincNet setup}

The SincNet model \cite{ravanelli2018speech} is built with a 4-layer CNN followed by a 5-layer DNN. The first layer of the CNN is constructed with trainable filters.
Those filters are initialised as a rectangular bandwidth mel-scale filterbank, an easily computable type of filter in the time domain.
Since the inverse Fourier transform of a rectangular low-pass filter is a sinc function, a rectangular bandwidth filter can be written as the difference of two low-pass filters as in equation \ref{eq:rect}.
\begin{equation}\label{eq:rect}
    h[n] = \sinc(2\pi f_2 n) - \sinc(2\pi f_1 n),
\end{equation}
where $h[n]$ represents a typical filter of the first convolutional layer.
The trainable parameters of the SincNet filters are the lower frequency  ($f_1$) and the bandwidth ($f_2 - f_1$), i.e., linear combinations of $f_1$ and $f_2$.
Moreover, in the time domain, filtering a signal is mathematically equivalent to the convolution of this signal with the filter kernel. 

Between the different convolution layers the following operations are used: maxpooling for downsampling, layernorm, ReLU and dropout before passing through a five-layer DNN. The input of signal is a raw waveform of 200 ms at 16 kHz. 
For this research the experiments are performed on TIMIT \cite{garofolo1993timit} and to verify that the observations are not database related, the baseline experiments have been double checked with the mini-Librispeech database \cite{panayotov2015librispeech}.

\subsection{Baseline}
\begin{figure}[]
    \centering
    \includegraphics[width = 1.05\columnwidth]{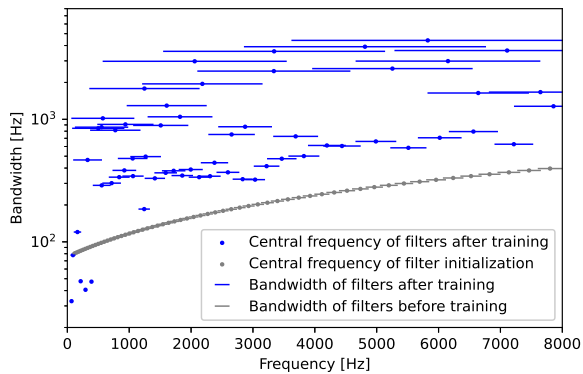}
    \caption{Evolution of the baseline implementation of SincNet: the graph on the top shows the initial filter distribution and the bottom plot show the filter distribution after training. The x-axis represents the frequency range and the y-axis the amplitude of the filters. The filters themselves are represented by their central frequency (dot) and their bandwidth (bar). }
    \label{fig:filter_evolution_128}
\end{figure}  
In the default implementation, the number of filters is initialised to 128 followed by 3 CNN layers of 60 filters each. 
The filter distribution for a similar experiment (60 filters on the first layer) is illustrated on figure \ref{fig:filter_evolution_128}.
We observe that some filters with a comparatively narrow bandwidth train towards a filterbank. The others train towards wider bandwidth filters. 
Concerning the frequency range, the wide-band filters could in principle be reconstructed with a linear combination of narrow band filters. In this paper, those two types of filters will be refered to as narrow and wide-band filters. The first part of this study focuses on the narrow band filters, since a large number of the wide-band filters seem to overfit the data.

\subsection{Number of filters}

Some filters in the first convolutional layer stay narrow-band while the others train towards wider bandwidths. Table \ref{tab:SincNetExp} gives an overview of the number of narrow band filters as well as the phone error rate (PER) on TIMIT.
To determine the number of narrow band filters an ad-hoc pruning operation has been applied after the filter training: the filters with wide bandwidths covering parts of the spectrum that are already covered by smaller bandwidth filters are discarded and only one filter is taken into account around the Nyquist frequency.

\begin{table}[t]
    \centering
    \caption{Filter pruning experiment: numbers of narrow band filters and related PER in function of the initialization.}
    \begin{tabular}{P{1.5cm}lllP{1.5cm}c}
    \toprule
    Sinc-Layer num. filters & \multicolumn{3}{l}{CNN-layers} & narrow band filters & PER (\%)  \\
    \midrule
    128        & 60       & 60       & 60       &  39  & 17.1\\
    100        & 60       & 60       & 60       &  45  & 17.1\\
    80         & 60       & 60       & 60       &  38  & 17.2\\
    60         & 60       & 60       & 60       &  32  & 17.4\\
    40         & 60       & 60       & 60       &  27  & 17.5\\
    30         & 60       & 60       & 60       &  24  & 17.5\\
    \bottomrule
    \end{tabular}
    \label{tab:SincNetExp}
\end{table}

The number of filters needed by the model to build a filterbank covering the whole frequency range can be determined by the number of narrow band filters. From table \ref{tab:SincNetExp} we can deduce that above 30 filters, the number of narrow-band filters that describe the frequency range is around 30 - 40 filters, this correlates with the results obtained by Zeghidour \cite{zeghidour2018learning} using Gabor filters.

We also notice that when the first layer is initialized to 30 or 40 filters (corresponding to the number of narrow-band filters of other layers), some of those filters still train toward larger band filters.
To analyse whether keeping the initilization to the initial scale performs as well as the combination of narrow and wide-band filters that the model learns, experiments have been made on 30 and 40 filters for fixed and non-fixed filters, the results are given in table \ref{tab:SincNetExp2}.
This raises the hypothesis that the wide-band filters are bringing some information not provided by the narrow band filters.
\begin{table}[t]
\centering
\caption{SincNet experiment: compare the performance of the training with the filters fixed and the filters that are free to train.}
\begin{tabular}{c|P{1cm}P{1cm}|P{1cm}P{1cm}}
\toprule
            & \multicolumn{2}{P{2.5cm}|}{fixed filters} & \multicolumn{2}{P{2.5cm}}{trained filters} \\
            
num. filters & loss            & PER            & loss             & PER              \\
\midrule
40          & 2.35            & 18.3           & 2.31             & 17.6             \\
30          & 2.37            & 18.0           & 2.33             & 17.5         \\
\bottomrule
\end{tabular}
\label{tab:SincNetExp2}.
\end{table}

\subsection{Scale after training}
Given that there are several frequency warpings that can be justified from a physiological point of view, it is informative (and simple) to investigate which one is preferred by an E2E system.  It is clear by inspection that it is the narrow band filters that learn the warped filterbank.  In \cite{agrawal2019unsupervised} a convolutional variational autoencoder (CVAE) architecture that learns convolutional filters from raw waveforms using unsupervised learning was proposed.  However, the analysis was only based on the central frequencies learned by those filters, not the narrow/wide-band distinction. In the present paper the central frequencies of only the narrow band filters are taken into account.

\subsubsection{Experiment}

The experiment consisted simply of analysing which filterbank the narrow-band filters trained above were learning.  The experiment was repeated for several models with different initializations.
The metric used to compute the distance between the initial and trained scale is the mean of the Euclidean distance:
\begin{equation}
    d(x, s) = \frac{1}{N}\sqrt{ \sum_{i=0}^N (x_i - s_i)^2}
\end{equation}

\begin{table}[t]
\caption{Mean Euclidean distance between narrow bandfilter's normalized central frequency distribution and different scalings for different amount of filters (Mel filterbank) and different initial scalings (30 filters).}
\begin{tabular}{l|llll}
\toprule
Initialized to  & \multicolumn{4}{l}{Compared to}   \\
scale - filters  & Mel   & Bark  & ERB   & Greenwood \\
\midrule

Mel - 128     & \textbf{0.0023} & 0.0047 & 0.0070 & 0.0086 \\

Mel - 60      & \textbf{0.0018} & 0.0044 & 0.0070 & 0.0088 \\
Mel - 40      & \textbf{0.0022} & 0.0039 & 0.0065 & 0.0084 \\
Mel - 30      & \textbf{0.0020} & 0.0043 & 0.0071 & 0.0091 \\
Bark - 30     & \textbf{0.0025} & 0.0037 & 0.0062 & 0.0082 \\
ERB  - 30     & 0.0030 & \textbf{0.0029} & 0.0055 & 0.0076 \\
Greenwood -30 & \textbf{0.0037} & 0.0068 & 0.0095 & 0.0116 \\
\bottomrule
\end{tabular}
\end{table}

The narrow band filters of a filterbank initialized to the mel scale remain mel-scale distributed. 
When initializing 30 -- 40 filters as starting point to different scalings, other scalings also train towards mel scale.  It follows that the mel scaling is an appropriate choice for filterbank initializations.

\subsection{Corollary}
The results so far have reinforced that E2E approaches do indeed learn what has been known for many years about cochlear models: that 30 to 40 filters are sufficient and that, regardless of physical measurements, the mel scale is the one that is perceptually important.
However, from figure \ref{fig:filter_evolution_128} it is clear that SincNet filters train towards a mixture of narrow \emph{and wide-band} filters. Moreover from table \ref{tab:SincNetExp}, in all the experiments done for this section, the model always learns wide-band filters. It follows that these wide-band structures are important.
Two questions arise:
\begin{enumerate}
    \item Can the filters be initialized to wide-band, removing or reducing the need to train them?
    \item Why do wide-band filters appear at all given that they are, to a first approximation, just linear combinations of narrow-band filters? 
\end{enumerate}
These are addressed in the following section.

%% file: Wideband_filters.tex
\section{Wide-band filter analysis}

\subsection{Wide-band initialization}
\begin{figure}
    \centering
    \includegraphics[width = 0.95\columnwidth]{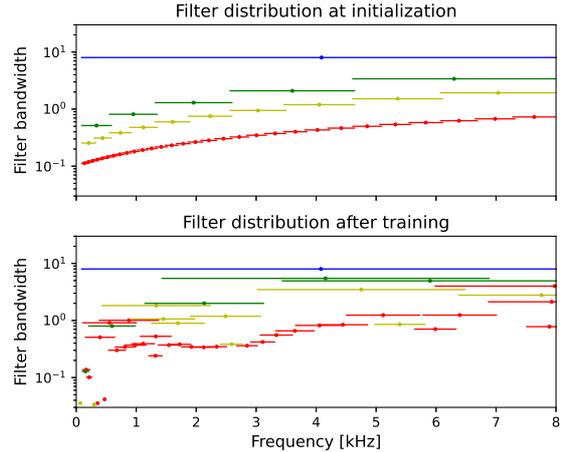}
    \caption{Filter repartition of superimposed filterbanks before (top plot) and after (bottom plot) training. In the initialization, the red filterbank is a narrow-band filterbank composed of 30 filters. The rest are filterbanks of 10, 5 and 1 filters capturing information that could in principle be reconstructed by a combination of the narrow band filters.}
    \label{fig:new_init}
\end{figure}
\subsubsection{Hypothesis}
Wide-band filters are important; it follows that the training can be optimized by initializing a narrow band filterbank as before and adding wide-band filters in addition of those filters.
This hypothesis can be confirmed by an experiment comprising initializing several frozen superimposed filterbanks where the wide-band filters are combinations of several narrow band filters.

\subsubsection{Experiment}
Figure \ref{fig:new_init} shows the initialization and training of a model built with 4 ranges of filters illustrated on the upper plot.
  
An estimation of the filter distribution after training of this new initialisation is illustrated in the bottom plot.
Four experiments using those filters are summarized in table \ref{tab:wb}:
\begin{table}[t]
\caption{Summary of experiments using narrow and/or wide-band filters}
\begin{tabular}{lllll}
\toprule
filters         & filter type                 & fixed & loss  & PER  \\ 
\midrule
10 - 5 - 1      & wide           & yes           & 2.410 & 18.6\% \\
30              & narrow         & yes           & 2.374 & 18.0\%   \\
30 - 10 - 5 - 1 & narrow \& wide & yes           & 2.335 & 17.7\%   \\
30 - 10 - 5 - 1 & narrow \& wide & no            & 2.306 & \textbf{17.5}\%  \\
\bottomrule
\end{tabular}
\label{tab:wb}
\end{table}

Using only the narrow band filters gives a final PER of 18\%, the combination of narrow and wide-band filters give for frozen filters a PER of 17.7\% and for trained filters a PER of 17.5\%. A combination of narrow and wide-band frozen filters already gives an improvement of 0.3\% WER. The same effect is observed on the loss: the loss for a combination of narrow and wide-band filters is lower than for only narrow band filters.
The new initialization is consequently closer to what the model learns compared to the baseline initialization. Aside, it is interesting to notice from figure \ref{fig:new_init} that when trained most of the narrow band filters stay narrow band and most of the wide-band filters stay wide-band.

Thus the hypothesis is demonstrated.  We conclude that it can indeed be beneficial to provide a mixture of narrow and wide-band filters in an ASR front-end.

\subsection{Why wide-band filters?}


Wide-band filters are in principle linear combinations of several narrow band filters; the network should be able to learn such a combination trivially, much as we assume it is learning the energy feature that was commonly used in HMM-based models.  The most plausible explanation for the network's failure to do so arises from the interaction of harmonic (voiced) and aperiodic (unvoiced) components.  Harmonic components in the same filter add constructively in the magnitude domain.  Aperiodic components, however, add as random variables; the variances add leading to a magnitude reduction by a factor of $\sqrt{N}$ for $N$ discrete components.  The wider band filters hence tend to favour the voiced components.

In complex narrowband (e.g., Fourier transform based) filters, the squaring operation leads to both harmonic and aperiodic components adding \emph{in the same ratio} in the magnitude or power domain, inhibiting emphasis of the harmonic component.  \sincnet comprises real-valued filters; however, the subsequent network architecture can inhibit the behaviour.  Each convolutional layer is followed by four typical operations: max-pooling (downsampling), layernorm, ReLU (activation function) and dropout that have some influence on the signal.  Of these, the maxpooling function and ReLU activation bring some distortions to the signal.

\subsubsection{Hypothesis}

It is possible to design a simple experiment to examine whether the above non-linearities inhibit simulation of wideband filters.  The experiment encompasses three intuitive hypotheses:\\

    1. \textbf{Using average-pooling instead of maxpooling} removes the noise artifacts that are created by maxpooling on the filtered signal, but since we continue to use a pooling function, aliasing still happens for the high frequencies. 
    In \cite{dubey2019transfer} some experiments showed that using average pooling reduced the WER but without explaining the possible reasons.\\
    2. \textbf{ Moving the first downsampling factor towards a further layer} inhibits downsampling just after the filtering of the signal, this removes both the effect of aliasing and signal distortion (although it increases the data size at several convolution layers).\\
    3. By inspection, \textbf{using a tanh or sigmoid function} removes some low frequency artifacts created by the ReLU function.  However, it is well known that the cochlea contains a rectifier function \cite{deBoer1978}, implying that ReLU is the right physiologically plausible solution.  It is not clear a-priori which of these properties is more important.

\subsubsection{Experiment}

Table \ref{tab:analysis} summarises the performance of the experiments implied above after the first convolution layer: use average pooling, move the first downsampling factor to a later layer and check that ReLU is appropriate.
\begin{table}[t]
\caption{The effects of modifying the downsampling and pooling schemes. The numbers in the second column refer to the downsampling rate at each of the pooling operations in the convolutional layers (1 implies no downsampling).}
\begin{tabular}{llP{1,3cm}ll}
\toprule
filters & downsampling & pooling 1st layer & act. function & PER    \\ 
\midrule
30      & 3-3-3-2      & max                     & ReLU                & 17.5\% \\
30      & 3-3-3-2      & avg                     & ReLU                & 17.1\% \\
30      & 1-3-3-6      & -                       & ReLU                & \textbf{16.8}\% \\
30      & 1-3-3-6      & -                       & sigmoid             & 18.1\%\\
\bottomrule
\end{tabular}
\label{tab:analysis}
\end{table}

Replacing the max-pooling with average-pooling leads to an improvement in performance.  Displacing the downsampling by one layer, in principle allowing the network to combine filters, leads to a further improvement.  This broadly demonstrates the first two points of our hypothesis.
Changing the activation function to sigmoid deteriorates the PER.  This tends to confirm that the physiologically-implied rectifier --- yielding a simple spectral envelope --- is also the right solution in the artificial solution.
We note that, even with the best performing architecture, the system still learns some wide-band filters.  This implies that our solution is not perfect.